\documentclass[sigconf,authorversion]{acmart}

\newcommand{\acmtog}{1}

\newcommand{\final}{1}
\newcommand{\proteus}{1}
\newcommand{\samp}{1}

\newcommand{\authorversion}{1}
\newcommand{\usecleveref}{1}

\usepackage{color}
\usepackage{ifthen}
\usepackage{float}
\usepackage{alltt}
\usepackage{amsthm}
\usepackage{newlfont} 
\usepackage{fixltx2e}
\usepackage{subfig} 
\usepackage{multirow}
\ifdefined\usecleveref
\usepackage{cleveref} 
\fi
\usepackage{algorithmic}
\usepackage{mathtools} 
\usepackage[ruled]{algorithm2e} 

\ifdefined\acmtog
\setcitestyle{square}
\def\textrightarrow{$\rightarrow$}
\fi

\ifdefined\qed
\else
\newcommand{\qed}{\hfill\blacksquare}
\fi

\newcommand{\pcomment}[1]{\mbox{// {\it #1}}}  

\newcommand{\captionfontsize}{8.7pt}
\newcommand{\captionfontskip}{10.4pt} 
\ifdefined\siggraph
\LetLtxMacro\oldcaption\caption
\renewcommand{\caption}[2][]{\oldcaption[#1]{{\fontsize{\captionfontsize}{\captionfontskip}\selectfont #1} {\fontsize{\captionfontsize}{\captionfontskip}\selectfont #2}}}
\newcommand{\Caption}[2]{\caption[#1]{#2}}
\else
\SetAlFnt{\small}
\SetAlCapFnt{\small}
\SetAlCapNameFnt{\small}
\SetAlCapHSkip{0pt}
\IncMargin{-\parindent}
\newcommand{\Caption}[2]{\caption[#1]{{\em #1} #2}}
\fi

\newcommand{\emacsquote}[1]{{``#1''}}

\definecolor{SithColor}{rgb}{0.7,0,0} 
\newcommand{\liyi}[1]{{\color{SithColor} Li-Yi: #1 $\qed$}}
\definecolor{ConsularColor}{rgb}{0,0.4,0} 
\definecolor{GuardianColor}{rgb}{0,0,0.8} 
\definecolor{WinduColor}{rgb}{0.56,0.34,0.62} 
\definecolor{ImplementColor}{RGB}{169,52,31} 
\newcommand{\chsu}[1]{{\color{GuardianColor} Chen-Yuan: #1 $\qed$}} 
\newcommand{\liyinote}[1]{{\color{WinduColor} Li-Yi: #1 $\qed$}} 
\newcommand{\implemento}[1]{{\color{ImplementColor}#1}} 
\newcommand{\shally}[1]{{\color{ConsularColor} Shally: #1 $\qed$}} 
\newcommand{\warning}[1]{{\it\color{red} #1}}
\newcommand{\note}[1]{{\it\color{ConsularColor} #1}}
\newcommand{\nothing}[1]{}

\definecolor{AudioColor}{rgb}{0.56,0.34,0.62}
\newcommand{\audio}[2][]{{\color{AudioColor} Audio: #2 $\qed$}}
\definecolor{VideoColor}{rgb}{0.44,0.66,0.38}

\definecolor{TodoColor}{rgb}{1, 0, 0} 

\definecolor{DoneColor}{rgb}{0.1,0.6,1.0} 

\definecolor{DeadlineColor}{rgb}{0.9,0.4,0} 

\definecolor{figred}{rgb}{1,0,0}
\definecolor{figgreen}{rgb}{0,0.6,0}
\definecolor{figblue}{rgb}{0,0,1}
\definecolor{figpink}{rgb}{1,0.63,0.63}

\ifdefined\usecleveref
\newcommand{\figref}[1]{\Cref{#1}}
\newcommand{\algref}[1]{\Cref{#1}}
\newcommand{\secref}[1]{\Cref{#1}}
\newcommand{\eqnref}[1]{\Cref{#1}}
\else
\newcommand{\figref}[1]{Figure~\ref{#1}}
\newcommand{\algref}[1]{Algorithm~\ref{#1}}
\newcommand{\secref}[1]{Section~\ref{#1}}
\newcommand{\eqnref}[1]{Equation~\ref{#1}}
\fi

\ifthenelse{\equal{\final}{1}}
{
\renewcommand{\liyi}[1]{}
\renewcommand{\chsu}[1]{}
\renewcommand{\shally}[1]{}
\renewcommand{\warning}[1]{}
\renewcommand{\note}[1]{}
\renewcommand{\liyinote}[1]{}
}
{}

\ifdefined\proteus
\else
\renewcommand{\implemento}[1]{}
\fi

\ifdefined\sigchi

\else

\fi

\newcommand{\pseudocode}{Algorithm}
\floatstyle{plain}
\newfloat{algorithm}{tbhp}{lop}
\floatname{algorithm}{\pseudocode}

\newcommand{\filename}[1]{\url{#1}}
\newcommand{\foldername}[1]{\url{#1}}
\ifdefined\email
\else
\newcommand{\email}[1]{\url{#1}}
\fi

\let\oldparagraph\paragraph
\iffalse
\newcommand{\passage}[1]{\oldparagraph{\textbf{#1}}}
\renewcommand{\paragraph}[1]{\oldparagraph{\textbf{#1}.}} 
\else
\newcommand{\passage}[1]{\oldparagraph{{#1}}}
\renewcommand{\paragraph}[1]{\oldparagraph{{#1}\ifdefined\sigchi\else.\fi}}
\fi

\newcommand{\point}{\mathbf{p}}

\newcommand{\sample}{s}

\newcommand{\element}{e}

\ifdefined\position
\renewcommand{\position}{\point}
\else
\newcommand{\position}{\point}
\fi

\newcommand{\weight}{\mathbf{w}}

\newcommand{\weightField}{\mathbf{w}}

\newcommand{\weightSample}{\weightField_s}

\newcommand{\assign}{\leftarrow}

\newcommand{\sampleSet}{\mathbf{S}}
\newcommand{\norm}[1]{\left\lVert #1 \right\rVert}

\newcommand{\spacingRatio}{\xi}
\newcommand{\satTree}{\mathbf{T}}
\newcommand{\sampleBall}{b}
\newcommand{\sampleBallPrime}{\sampleBall^\prime}
\newcommand{\includePolarBall}{include}

\newcommand{\elementPoint}{p}
\newcommand{\elementPointSet}{\mathbf{P}}
\newcommand{\samplePointSet}{\mathbf{P}}
\newcommand{\sampleBallNumber}{B}

\ifdefined\usemathtools

\DeclarePairedDelimiter\floor{\lfloor}{\rfloor}
\DeclarePairedDelimiter\abs{\lvert}{\rvert}
\else

\newcommand{\floor}[1]{\lfloor #1 \rfloor}
\newcommand{\abs}[1]{\lvert #1 \rvert}
\fi

\newcommand{\radius}{r}
\newcommand{\elementArea}{D_A}
\newcommand{\radiusMin}{\radius_{min}}
\newcommand{\sampleAreaMin}{S_{A_{min}}}
\newcommand{\sampleCount}{N}
\newcommand{\sampleCountMin}{\sampleCount_{min}}
\newcommand{\sampleCountMax}{\sampleCount_{max}}
\newcommand{\cost}{C}
\newcommand{\measure}{M}
\newcommand{\overlap}{o}
\newcommand{\exterior}{e}
\newcommand{\adjacency}{a}
\newcommand{\asymmetry}{y}
\newcommand{\measureOverlap}{\measure_{\overlap}}
\newcommand{\measureExterior}{\measure_{\exterior}}
\newcommand{\measureAdjacency}{\measure_{\adjacency}}
\newcommand{\measureAsymmetry}{\measure_{\asymmetry}}
\newcommand{\weightOverlap}{\weight_{\overlap}}
\newcommand{\weightExterior}{\weight_{\exterior}}
\newcommand{\weightAdjacency}{\weight_{\adjacency}}
\newcommand{\weightAsymmetry}{\weight_{\asymmetry}}
\newcommand{\adjacencyTotal}{A}
\newcommand{\adjacencyTotalMax}{\adjacencyTotal_{max}}
\newcommand{\adjacencyTotalMin}{\adjacencyTotal_{min}}
\newcommand{\pixelCount}{p}
\newcommand{\quadrantArea}{Q}
\newcommand{\mul}{}

\graphicspath{
{./figs/vector/}
{./figs/raster/}
}

\def\plaintitle{Simple Methods to Represent Shapes with Sample Spheres}
\def\plainshorttitle{Simply Sample Shapes with Spheres}
\def\plainauthor{Li-Yi Wei}
\def\plainaffiliation{Adobe Research}
\def\emptyauthor{}
\def\plainkeywords{element, sampling}

\def\adoberesearch{\plainaffiliation}
\def\adobeinc{Adobe Inc.}

\ifdefined\sigchi
\ifdefined\nocopyright
\else
\CopyrightYear{2020}
\setcopyright{acmcopyright}
\toappear{\scriptsize
\ifdefined\authorversion
This is the author's version of the work.
It is posted here for your personal use. Not for redistribution.
The definitive Version of Record was published in\\
\else
Permission to make digital or hard copies of all or part of this work for personal or classroom use is granted without fee provided that copies are not made or distributed for profit or commercial advantage and that copies bear this notice and the full citation on the first page. Copyrights for components of this work owned by others than ACM must be honored. Abstracting with credit is permitted. To copy otherwise, or republish, to post on servers or to redistribute to lists, requires prior specific permission and/or a fee. Request permissions from permissions@acm.org. \\
\fi
{\emph{CHI '20, April 25--30, 2020, Honolulu, HI, USA.} } \\
\copyright~2020 Association for Computing Machinery. \\
ACM ISBN 978-1-4503-6708-0/20/04\ ...\$15.00.\\
http://dx.doi.org/10.1145/3313831.3376248
}
\doi{https://doi.org/10.1145/3313831.XXXXXXX}
\isbn{978-1-4503-6708-0/20/04}
\conferenceinfo{CHI'20,}{April  25--30, 2020, Honolulu, HI, USA}
\acmPrice{\$15.00}
\fi

\makeatletter
\def\url@leostyle{%
  \@ifundefined{selectfont}{
    \def\UrlFont{\sf}
  }{
    \def\UrlFont{\small\bf\ttfamily}
  }}
\makeatother
\def\pprw{8.5in}
\def\pprh{11in}

\newlength\savedintextsep
\newlength\savedcolumnsep
\definecolor{linkColor}{RGB}{6,125,233}
\hypersetup{%
  pdftitle={\plaintitle},
  pdfauthor={\emptyauthor},
  pdfkeywords={\plainkeywords},
  pdfdisplaydoctitle=true, 
  bookmarksnumbered,
  pdfstartview={FitH},
  colorlinks,
  citecolor=black,
  filecolor=black,
  linkcolor=black,
  urlcolor=linkColor,
  breaklinks=true,
  hypertexnames=false
}

\fi
\ifdefined\acmtog
\copyrightyear{2020}
\acmYear{2020}
\setcopyright{acmlicensed}\acmConference[SA '20 Technical Communications]{SIGGRAPH Asia 2020 Technical Communications}{December 4--13, 2020}{Virtual Event, Republic of Korea}
\acmBooktitle{SIGGRAPH Asia 2020 Technical Communications (SA '20 Technical Communications), December 4--13, 2020, Virtual Event, Republic of Korea}
\acmPrice{15.00}
\acmDOI{10.1145/3410700.3425424}
\acmISBN{978-1-4503-8080-5/20/11}
\acmSubmissionID{tcom\_108} 

\fi

\begin{document}

\title{\plaintitle}

\ifdefined\acmtog

\renewcommand{\shorttitle}{\plainshorttitle}

\author{Li-Yi Wei}
\affiliation{\institution{\adoberesearch}}

\author{Arjun V Anand}
\affiliation{\institution{\adobeinc}}

\author{Shally Kumar}
\affiliation{\institution{\adobeinc}}

\author{Tarun Beri}
\affiliation{\institution{\adobeinc}}

\renewcommand{\shortauthors}{Wei, Anand, Kumar, and Beri}

\else
\author{\plainauthor}
\affiliation{\institution{\plainaffiliation}}
\fi

\ifdefined\acmtog
\setcopyright{acmlicensed}
\fi

\ifdefined\sigchi
\maketitle
\fi

\begin{abstract}

Representing complex shapes with simple primitives in high accuracy is important for a variety of applications in computer graphics and geometry processing.
Existing solutions may produce suboptimal samples or are complex to implement.
We present methods to approximate given shapes with user-tunable number of spheres to balance between accuracy and simplicity: touching medial/scale-axis polar balls and k-means smallest enclosing circles.
Our methods are easy to implement, run efficiently, and can approach quality similar to manual construction.

\end{abstract}

%
%


%
%
\newcommand{\keywordlist}{\plainkeywords}

\ifdefined\acmtog
\fi

\ifdefined\sigchi
\fi

\ifdefined\acmtog
\maketitle
\fi

\newcommand{\sampmethod}{1}
\section{Background}

Natural and man-made objects come into a variety of sizes and shapes.
These geometry shapes play a central role in many scientific and engineering disciplines.
Some shapes can be complex and thus challenging to store and compute.
Thus, how to represent general shapes in sufficiently simple and yet accurate forms is an important research question.

Ideally, we would like to represent a given shape (e.g., a polygonal mesh model or a vector graphics design) using a small number of simple primitives (e.g., spheres) that can approximate the desired \nothing{numerical }properties of the original shapes with sufficient accuracy.
The definition of the latter is often application dependent, such as collision handling for physics simulation \cite{Hubbard:1996:APS}, level of details for progressive rendering \cite{Rusinkiewicz:2000:QMP}, shadow computation for global illumination \cite{Wang:2006:VSS}, control handles for deformation \cite{Jacobson:2011:BBW,Phogat:2019:SVG}, or element distribution for design \cite{Hsu:2020:AEF}.

\begin{figure}[htb!p]
\vspace{-1em}
\hspace{-1.2ex}
  \subfloat[grass]{
    \label{fig:element_representation:grass}
\ifdefined\brief		
    \includegraphics[width=0.235\linewidth]{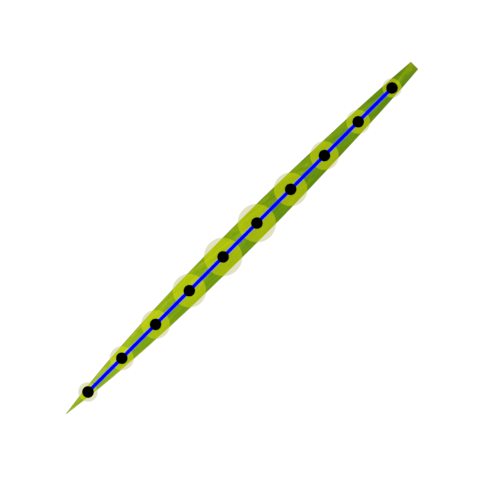}
\else
    \includegraphics[width=0.235\linewidth]{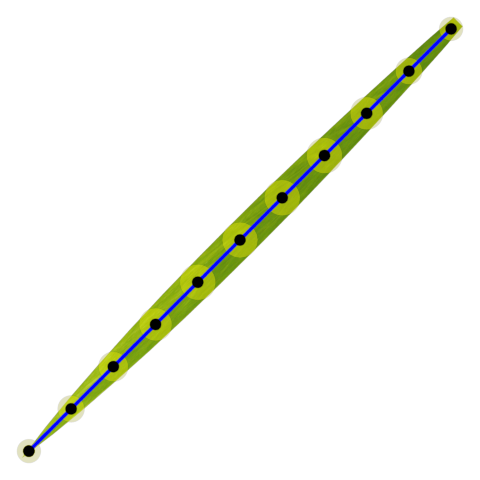}
\fi
  }
\hspace{-1.4ex}
  \subfloat[leaf]{
    \label{fig:element_representation:leaf}
    \includegraphics[width=0.235\linewidth]{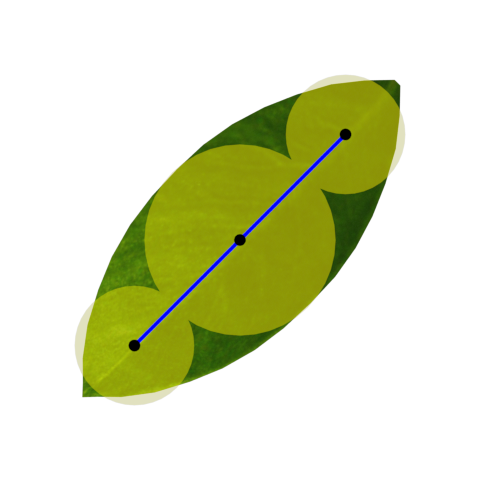}
  }
\ifdefined\brief  
  \subfloat[eggplant]{
    \label{fig:element_representation:eggplant}
    \includegraphics[width=0.235\linewidth]{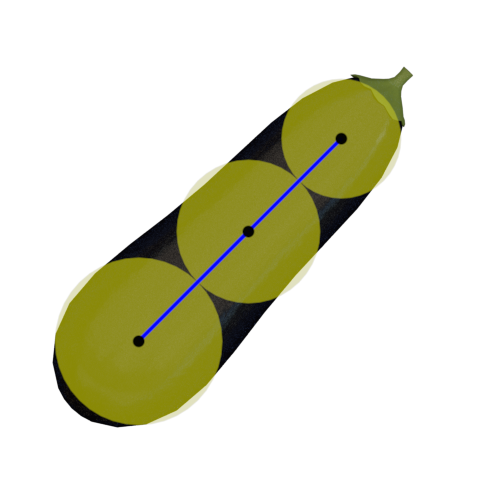}
  }
\else
\hspace{-1.4ex}
  \subfloat[semiquaver]{
    \label{fig:element_representation:semiquaver}
	\includegraphics[width=0.235\linewidth]{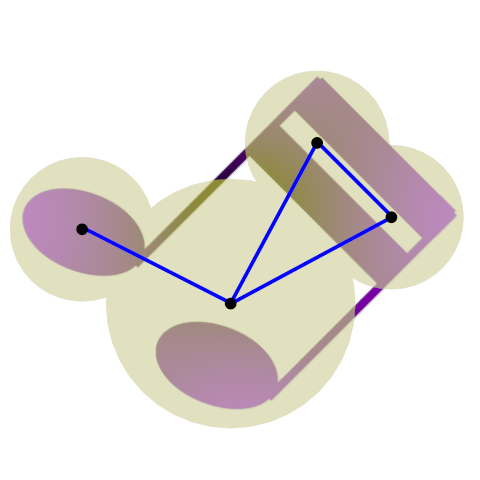}
  }
\fi   
\hspace{-1.4ex}
  \subfloat[treble clef]{
    \label{fig:element_representation:treble}
	\includegraphics[width=0.235\linewidth]{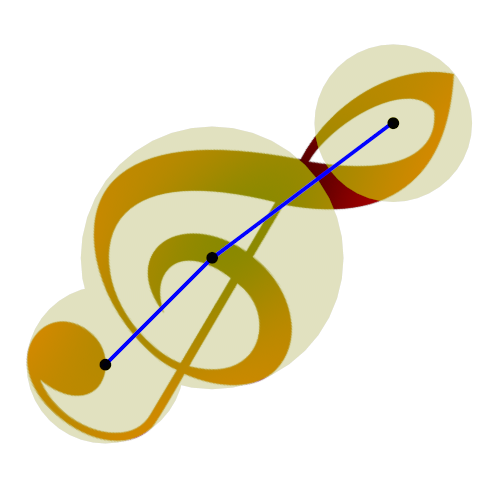}
  }
  \Caption{Element representation.}
  {%
  The black dots indicate the sample positions.
  The radii of the yellow circles represent the sample weights $\weightSample$.
  The blue lines denote the element graphs.
\ifdefined\brief
\else
  The sampling can be sparse \subref{fig:element_representation:grass}, dense \subref{fig:element_representation:leaf}, overlapping \subref{fig:element_representation:semiquaver} or hybrid \subref{fig:element_representation:treble}.
\iffalse
\liyi{(January 4, 2020) Avoid the potentially misleading information about automatic algorithm.}
\chsu{(January 4, 2020) These automatic algorithms can still assist users in sampling elements. It depends on how to use them.}
These examples are manually constructed.
\nothing{
Our current implementation lets users manually place and connect the samples to reflect their design intentions.
}
\else
\ifdefined\samp
These samples are manually placed and connected in \cite{Hsu:2020:AEF}.
\else
These samples can be either manually placed by designers or assisted by algorithms \cite{Amenta:2001:PC,Wang:2006:VSS,Thiery:2013:SSA,Li:2015:QCM}.
\fi
\fi
\nothing{
\liyi{(May 10, 2019) All graphs shown are 1D curves. Do we want to show other structure, like a tree or a mesh?}
\chsu{(May 11, 2019) Pebbles have mesh graph structures but they are not suitable for the example, and the remaining elements are mainly linear.
If necessary, I can modify one of the elements I used after.}
\liyi{(May 15, 2019) Can our algorithm handle general graphs not just linear ones? If so, we should show some examples as otherwise readers can misunderstand the scope of our method.}
}
\fi
  }
  \label{fig:element_representation}
\end{figure}

\begin{figure}[htb!p]
  \centering
  \includegraphics[width=\linewidth]{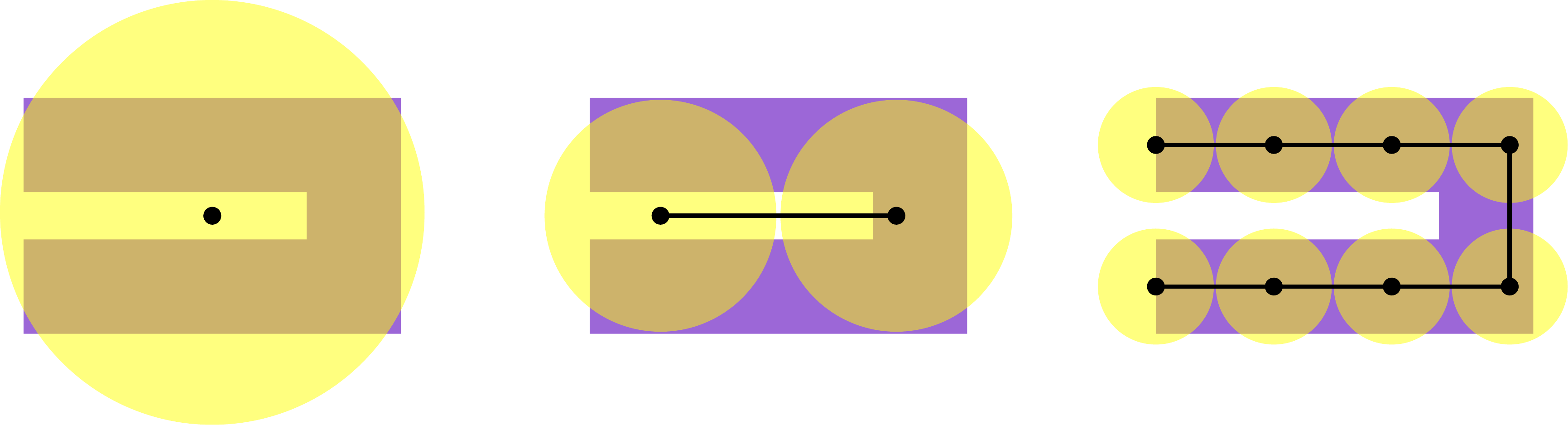}
  \Caption{Variety in element sampling.}
  {%
Depending on the design choice, the same element (purple) can be represented in different ways, such as 1 sample for a rigid component (left), 2 samples for a mildly deformable shape (middle), and 8 samples to allow further deformation among the two branches (right).
  }
  \label{fig:element_sample_variety}
\end{figure}

Due to the importance of this problem, a variety of solutions have been proposed to approximate shapes with sample spheres \cite{Wang:2006:VSS,Stolpner:2012:MSS,Thiery:2013:SSA}.
\nothing{
However, existing methods often require too many spheres or are too complex for implementation and efficient computation.
}
The method in \cite{Hsu:2020:AEF} lets users manually place and connect these spheres to reflect their design intention as illustrated in \figref{fig:element_representation} and \figref{fig:element_sample_variety} that might not be divinable by automatic computation.
\iffalse

As an automatic default, we can consider using the control points of the vector graphics representation of the elements as the samples, and set $\weightSample(\sample)$ as the half-way distance to the nearest neighbor.
\liyi{(January 4, check to make sure the control points are not too far away from the actual element shapes.}
The graph can be simplified via first performing a $k$-means clustering of the samples with $k$ a control parameter to trade off fidelity and complexity.
Another possibility is to compute a smooth \emacsquote{skin} of the element as in \cite{Saputra:2018:RPD,Saputra:2019:IDD}, and sample its medial axis with adjacent/touching polar balls \cite{Amenta:2001:PC}.
A practical implementation for vector graphics is via scale-axis transform from SVG \cite{Steenkamp:2019:MSA}.

\else
However, it might not be reasonable to require all users to go through this process as it is more natural for them to directly design the element shapes without thinking about the underlying implementation.
\ifdefined\samp
\else
\cite{Wei:2020:SMR} provide automatic methods to sample given element shapes with spheres.
\fi
\fi

The Puppet Warp tool in Adobe Illustrator has a machine-learning-based method for automatic pin placement \cite{Phogat:2019:SVG}, but it often produces undesirable outcomes, such missing symmetries and leaving one part of the shape inconsistent with the other (\figref{fig:comparison_pw_our}).

\nothing{
\cite{Alt:2006:MCP}

in computer graphics and geometry processing
}

\section{Method}
\label{sec:method}

We provide two automatic methods to sample given shapes with spheres.
Both methods are very simple to implement, run fast, and can well represent the original shapes with a few spheres, whose numbers can be tuned by the users or the rest of the authoring system as a quality-performance trade off:

\begin{algorithm}[tb]
  \begin{flushleft}

  $\spacingRatio \assign$ user specified spacing ratio $\in [0 \; \infty)$

  compute MAT/SAT $\satTree$ of a given element $\element$
  \pcomment{\cite{Amenta:2001:PC,Steenkamp:2019:MSA}}
  
  sample set $\sampleSet \assign \emptyset$
 
  \For{each polar ball $\sampleBall \in \satTree$ in decreasing radius}
  {
    $\includePolarBall \assign true$

    \For{each $\sampleBallPrime \in \sampleSet$}
    {
      \If{$\norm{\sampleBall_{center} - \sampleBallPrime_{center}} < \spacingRatio \left(\sampleBall_{radius} + \sampleBallPrime_{radius} \right)$}
      {
         $\includePolarBall \assign false$
      }
    }

    \If{$\includePolarBall$}
    {
      $\sampleSet \assign \sampleSet + \sampleBall$
    }
  }

  \Return{$\sampleSet$}
 
  \end{flushleft}
  \Caption{Element sampling via touching polar balls from a MAT/SAT tree.}
  {%
  }
  \label{alg:element_sample_mat}
\end{algorithm}

\begin{figure}[tbh!p]
  \centering
  \subfloat[semi breve]{
    \includegraphics[width=0.24\linewidth]{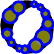}
  }
  \subfloat[minim]{
    \includegraphics[width=0.24\linewidth]{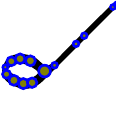}
  }
  \subfloat[semi quaver]{
    \includegraphics[width=0.24\linewidth]{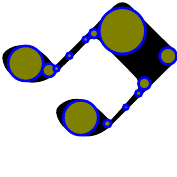}
  }
  \subfloat[treble clef]{
    \includegraphics[width=0.24\linewidth]{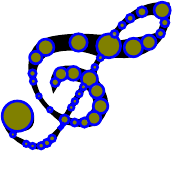}
  }
  \Caption{Automatic element sampling results via touching MAT/SAT polar balls.}
  {%
Users can set an upper limit on the overlap among sample spheres $\spacingRatio$ (for performance reason), and our method can optimize the number of samples, their locations, and weights\nothing{ for each element}.
In this example, $\spacingRatio = 0.8$.
  }
  \label{fig:element_sample_mat}
\end{figure}

\begin{algorithm}[tb]
  \begin{flushleft}

  $\elementPointSet(\element) \assign$ uniform point sets from a given element $\element$ 

  $\sampleBallNumber \assign$ user-specified number of samples for $\element$

  sample set $\sampleSet \assign$ random $\sampleBallNumber$ centers from $\elementPointSet(\element)$ with $0$ radii
 
  \While{not enough iterations}
  {
    \pcomment{compute nearest ball center for each point}

    \For{each $\sampleBall \in \sampleSet$}
    {
      $\samplePointSet(\sampleBall) \assign \emptyset$
    }

    \For{each $\elementPoint \in \elementPointSet(\element)$}
    {
      $\sampleBallPrime = argmin_{\sampleBall} \norm{\elementPoint - \sampleBall_{center}} , \; \forall \sampleBall \in \sampleSet$

      $\samplePointSet(\sampleBallPrime) \assign \samplePointSet(\sampleBallPrime) + \elementPoint$
    }
    
    \pcomment{compute smallest enclosing circles}

    \For{each $\sampleBall \in \sampleSet$}
    {
      $\sampleBall \assign SmallestEnclosingCircle\left( \samplePointSet(\sampleBall) \right)$
    }
  }

  \Return{$\sampleSet$} 
  \end{flushleft}
  \Caption{Element sampling via k-means smallest enclosing circles.}
  {%
  }
  \label{alg:element_sample_kmeans_sec}
\end{algorithm}

\begin{figure}[tbh!p]
  \centering
  \subfloat[semi breve]{
    \includegraphics[width=0.24\linewidth]{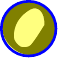}
  }
  \subfloat[minim]{
    \label{fig:element_sample_kmeans_sec:minim}
    \includegraphics[width=0.24\linewidth]{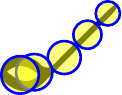}
  }
  \subfloat[semi quaver]{
    \includegraphics[width=0.24\linewidth]{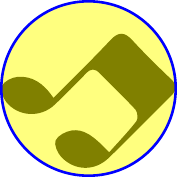}
  }
  \subfloat[treble clef]{
    \includegraphics[width=0.24\linewidth]{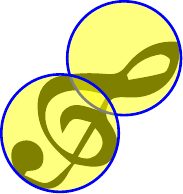}
  }
  \Caption{Automatic element sampling results via k-means smallest enclosing circle.}
  {%
Users can set an upper limit on the number of samples per element (for performance reason) $\sampleBallNumber$, and our method can optimize the number of samples, their locations, and weights for each element.
In this example, $\sampleBallNumber = 8$.
Notice the number of samples for different elements are automatically chosen depending on their shape complexity.
  }
  \label{fig:element_sample_kmeans_sec}
\end{figure}

\begin{figure}[tbh!p]
  \centering
  \subfloat[semi breve]{
    \includegraphics[width=0.24\linewidth]{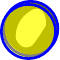}
  }
  \subfloat[minim]{
    \label{fig:element_sample_kmeans_sec_min2:minim}
    \includegraphics[width=0.24\linewidth]{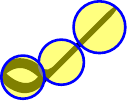}
  }
  \subfloat[semi quaver]{
    \includegraphics[width=0.24\linewidth]{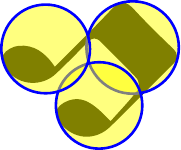}
  }
  \subfloat[treble clef]{
    \includegraphics[width=0.24\linewidth]{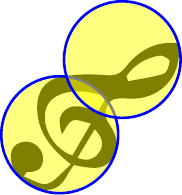}
  }
  \Caption{K-means smallest enclosing circle with at least two samples per element.}
  {%
  }
  \label{fig:element_sample_kmeans_sec_min2}
\end{figure}

\passage{Touching medial/scale axis polar balls (MAT)}

In this method, we sample the medial/scale axis polar balls from a given shape \cite{Amenta:2001:PC,Steenkamp:2019:MSA}, and select a subset in which any two spheres overlay no more than a user selected threshold $\spacingRatio$.
In our current implementation, we set a threshold of overlap between the radii of the two spheres.
For example, if the threshold is $0$, no spheres can intersect.
And if the threshold is $\infty$, all polar balls will be selected.
We find it helpful to select the polar balls in decreasing size so ensure the larger and thus more representative ones are retained first.
The method is summarized in \algref{alg:element_sample_mat}.
See \figref{fig:element_sample_mat} for examples.

\passage{K-means clustering with smallest enclosing circles (SEC)}
 
In this method, we combine k-means clustering with smallest enclosing circle \cite{Nayuki:2018:SEC,Wikipedia:2019:SCiP} to iterate the 2 steps: find the nearest ball center for each point on the element, and compute the smallest enclosing ball/circle from all belonging points (instead of the centroid step in k-means).
The points can be uniformly sampled from the element boundary paths (including interior holes).
The method is summarized in \algref{alg:element_sample_kmeans_sec}.
Users can specify a $[min, max]$ range of the number of samples for a given element, and run our algorithms to decide which number produces the minimal total sample ball areas.
\liyi{(June 22, 2020)
To encourage tighter fit for deformable elements, we can also consider minimal total sample/ball area outside the element shape.
}
\iffalse
\liyi{(June 27, 2020)
Another possibility is to start with a minimal number of samples higher than 1, to enforce deformability.
}
\else
($min > 1$ prevents 1-sample elements that can never deform.)
\fi
See \figref{fig:element_sample_kmeans_sec} and \figref{fig:element_sample_kmeans_sec_min2} for examples.

As can be seen, the MAT method tends to produce more spheres that all fit within the original shapes, while the SEC method tends to produce less spheres that cover the original shapes.
Thus, they are more suitable for deformable and rigid elements, respectively.

\paragraph{Graph construction}

After placing the samples, we can connect them to form element graphs (\figref{fig:element_representation} and \figref{fig:element_sample_variety}).
For a fully rigid shape, we can connect all pairs of samples to form a complete graph.
To allow (skeletal) deformation, we can connect along the medial/scale axis for MAT or overlapping/touching sample spheres for SEC.

\section{Refinement}

\begin{figure}[tbh!p]
  \centering
  \includegraphics[width=\linewidth]{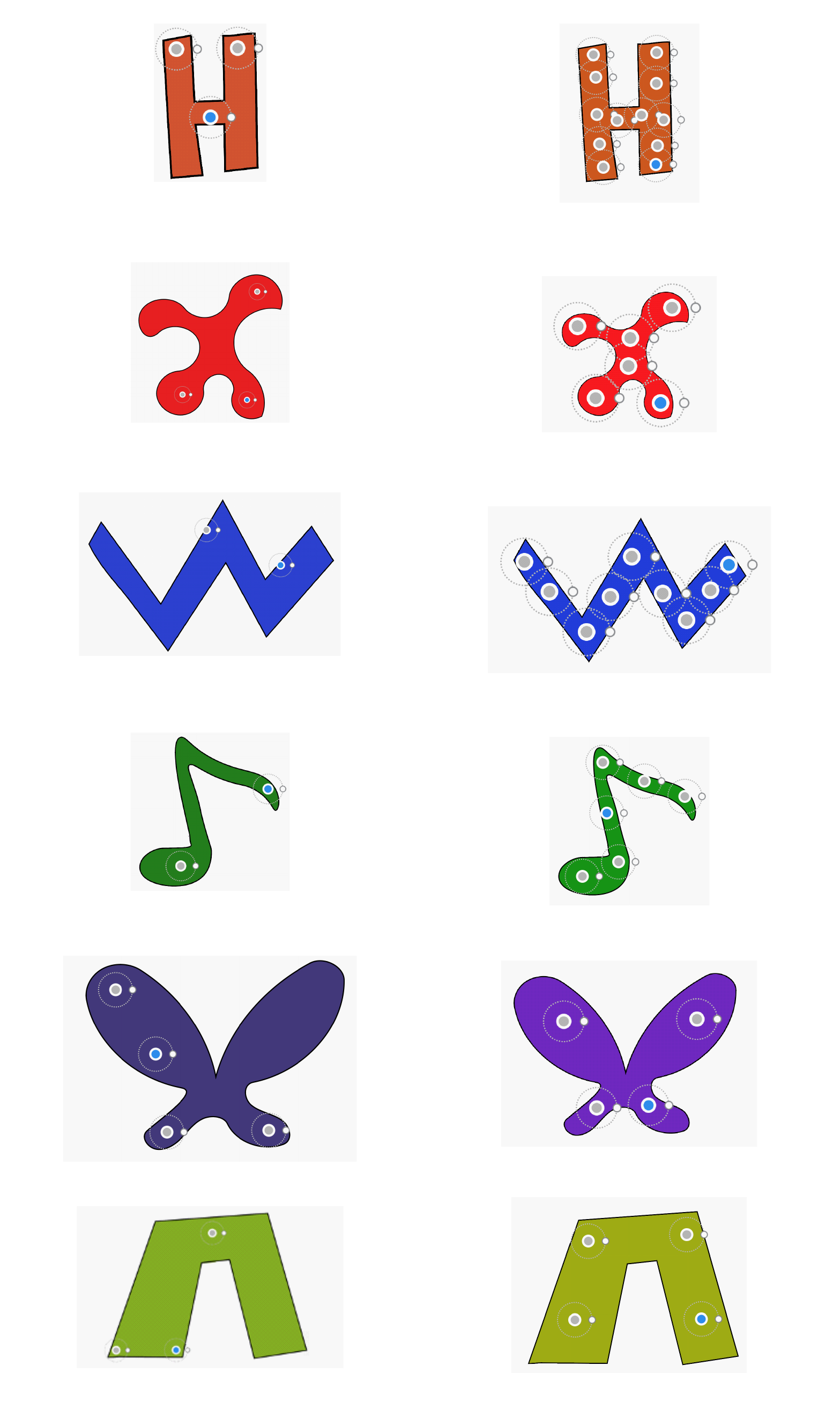}
  \begin{minipage}{0.48\linewidth}
    \centering
    \cite{Phogat:2019:SVG}
  \end{minipage}
  \begin{minipage}{0.48\linewidth}
    \centering
    Our SEC method
  \end{minipage}
  \Caption{Comparison with Adobe Illustrator Puppet Warp automatic pin placement \cite{Phogat:2019:SVG}.}
  {%
The results for \cite{Phogat:2019:SVG} were directly screen-captured and thus have different rendering effects from our results.
Notice that the samples produced via \cite{Phogat:2019:SVG} might miss key parts of the shapes, while our results can provide more balanced\nothing{comprehensive} coverage.
  }
  \label{fig:comparison_pw_our}
\end{figure}

\nothing{
This document explains how we can represent a 2D vector graphics shape as a collection of circles (henceforth, called {\em samples}). The proposed sampling is a pre-requisite for smooth automatic pattern completion through element fields. Additionally, the centers of the samples produced by our method can be directly used as {\em puppet warp} pins and this allows automatic generation of non-rigid deformable patterns inside a vector shape.
}

The SEC method in \secref{sec:method} can produce unpredictable results due to random initialization (e.g., comparing \figref{fig:element_sample_kmeans_sec:minim} and \figref{fig:element_sample_kmeans_sec_min2:minim}).
We describe a better initialization method and more principled criteria for choosing the number of samples.

\paragraph{Objectives}
\label{sec:Requirements}
The desired sampling\nothing{ (for {\em element fields} and {\em puppet warp})} should satisfy the following goals:
\begin{enumerate}
\item Sample centers should lie on, or close to, the medial axis of the input shape.
\item Overlap among samples should be minimal.
\item Samples should occupy minimum external area (i.e., area outside the input shape), but maximum internal area.
\item There should be a low number of adjacent samples.
One adjacent sample (for every sample) is okay, two are good if they lie on opposite sides, more than two are only desired in {\em 'T'} or '$+$' like shapes\nothing{ (at their joins)}.
\item Samples should symmetrically cover the shape, i.e., samples with left side outside the shape while right side inside the element are less preferred, while samples completely inside or equivalently outside on left/right and top/bottom are more preferred.
\end{enumerate}

\nothing{
Basically, we want to avoid the situations in \figref{fig:undesired} and get close to the desired sampling shown in \figref{fig:desired}.

\begin{figure}[tbh]
\begin{center}
\Caption{Undesired sampling (red lines represent medial axis).}
{}
\label{fig:undesired}
\end{center}
\end{figure}

\begin{figure}[tbh]
\begin{center}
\Caption{Desired sampling (red line represents medial axis).}
{}
\label{fig:desired}
\end{center}
\end{figure}
}

Some of the criterion mentioned \nothing{in \secref{sec:Requirements}}above might work against each other.
Let us consider the case of a narrow {\em U} shaped element versus a wider one.
If the parallel boundary walls of the element are far apart, then having a sample center at the medial axis would mean large samples.
This could increase the external area they occupy.
Thus, point $1$ might work against point $3$.
On the other hand, if we reduce the size of samples (and keep their centers at medial axis), it could reduce the external area but violate adjacencies.
Thus, we do not directly enforce keeping sample centers at medial axis and rather achieve that as a side-effect of other criteria.
We keep the external area low, maintain desired adjacency and symmetry to enforce sample centers to be as close to the medial axis as possible.

\paragraph{Initialization}

The {k-means} clustering in the SEC method requires a set of points to operate with.
Random initialization can produce unstable results; each time we execute sampling, a different result is returned.
To meet our first objective, we place points uniformly along the boundary and randomly in the interior of the input shape, and sort all the points first on y-axis and then on x-axis.
From this sorted point set, we pick up uniformly separated points (like rejection sampling for dart throwing \cite{Yuksel:2015:SEG}) for initialization.
This always produces stable results and has the tendency to converge faster.
\iffalse
Thus, we linearize our element and find points all along its boundary. We use standard Bezier approximation techniques for this. Additionally, we also create sufficient density of points inside the element.
\else
\fi


\paragraph{Optimizing sample count}

Let $\elementArea{}$ be the area of the input shape\nothing{ (to be sampled)} and $\radiusMin{}$ be the radius $\weightSample$ of the smallest sample that we would like to create.
If $\elementArea{}$ is less than the area of the smallest possible sample $\sampleAreaMin$ (i.e., $\pi \mul{} \radiusMin{}^{2}$), we treat that as a special case.
We can either rescale the shape\nothing{ (being vector)} or create one tight fitting sample (ignoring $\radiusMin{}$) located at the center of the shape.
Otherwise, we use the following algorithm to determine the number and locations of samples.

We first find the minimum and maximum possible number of samples.
Minimum count ($\sampleCountMin$) is default to 1 and maximum count ($\sampleCountMax$) is computed as $\floor{\frac{\elementArea{}}{\sampleAreaMin}}$.
We linearly scan all possibilities from $\sampleCountMin$ to $\sampleCountMax$ and find the number of samples that gives us the minimum cost, defined in \eqnref{eqn:cost}.
At every step (in the search space $\sampleCountMin$ to $\sampleCountMax$), we perform the SEC method (\algref{alg:element_sample_kmeans_sec}), and calculate the following energy value to identify the minimum that indicates the desired sample count:
\note{
\begin{figure*}[tb]
\begin{center}
\Caption{Results - left side is the input elements while right side is the output sampling.}
{}
\label{fig:results_no_normalization}
\end{center}
\end{figure*}
}
\begin{align}
\cost = \weightOverlap \mul{} \measureOverlap + \weightExterior \mul{} \measureExterior + \weightAdjacency \mul{} \measureAdjacency + \weightAsymmetry \mul{} \measureAsymmetry
\label{eqn:cost}
\end{align}
,
where the subscripts $\overlap$, $\exterior$, $\adjacency$ and $\asymmetry$ represent overlap, exterior area, adjacency, and asymmetry,
$\weight$ and $\measure$ represent weight and measure of the subscripted entities.
We normalize all measures on an equivalent scale $[0 \; 1]$ to avoid results being skewed towards any term. 

$\measureOverlap$ is computed as total overlapping area of all samples divided by total area of all samples. Since all samples are circles, this is straight forward to compute. Note that if two samples overlap, we count the overlapping area in both samples.

$\measureExterior$ is computed as the ratio of total area of samples not inside the shape to the total area of samples.
Note that we do not deal with overlaps here.
For every sample, we just compute how much of it is outside the shape.
This requires circle and triangle intersection evaluations which we perform via triangle tessellation of the input shape.

$\measureAdjacency$ is computed by finding the number of samples that are overlapping or adjoining.
If a sample has radius $\radius_1$ and another has radius $\radius_2$ and the separation between their centers is less than $\radius_1 + \radius_2$, we count that as an adjacency.
We also use some epsilon tolerance for stable results.
The extreme case is that every sample is adjacent to every other sample.
Thus, for $\sampleCount$ samples, the maximum possible number of adjacencies $\adjacencyTotalMax$ is given by $\sampleCount \mul{} (\sampleCount-1)$ and for a linearly connected sampling (like in a leaf shape in \figref{fig:element_representation:leaf}), the desired number of adjacencies $\adjacencyTotalMin$ is given by $2\mul{}(\sampleCount-1)$.
Thus, we can compute $\measureAdjacency$ as following:
\begin{align}
\measureAdjacency = \frac{\abs{\adjacencyTotal - \adjacencyTotalMin}}{\adjacencyTotalMax}
\end{align}
,
where $\adjacencyTotal$ is the total count of adjacencies among all samples.
Note that if two samples are adjacent, we count adjacency from both sides.
Since it will be rare to have $\adjacencyTotalMax$ adjacencies in a practical setup, there will be a higher tendency to have $\measureAdjacency$ less than $\measureOverlap$, $\measureExterior$ and $\measureAsymmetry$.
Thus, to counter this, we do two things: use $3\mul{}(\sampleCount-1)$ as a practical upper bound for $\adjacencyTotalMax$, and set $\weightAdjacency$ higher than $\weightOverlap$, $\weightExterior$ and $\weightAsymmetry$.
Also note that $\measureAdjacency$ is mostly immaterial (being zero) when number of samples is $2$.
Thus, in that scenario, we ignore $\measureAdjacency$ and evaluate cost $\cost$ by considering the other three measures only.

$\measureAsymmetry$ is computed by dividing every sample into four quadrants and measuring how much asymmetric coverage is exhibited by opposite quadrants (quadrant 1 versus 3 and quadrant 2 versus 4; quadrants are numbered anti-clockwise from x-axis).
We find the number of interior pixels $\pixelCount$ in each of the four quadrants and use the following formula to find $\measureAsymmetry$:
\begin{align}
\measureAsymmetry = \frac{0.5 \mul{} \abs{\pixelCount_1 - \pixelCount_3} + 0.5 \mul{} \abs{\pixelCount_2 - \pixelCount_4}}{Q}
\end{align}
,
where $\quadrantArea$ represents area of a quadrant; $\pixelCount_1$, $\pixelCount_2$, $\pixelCount_3$ and $\pixelCount_4$ represent interior pixels (i.e. inside the element) in quadrants 1, 2, 3 and 4.

\if false
Alternatively, we can do a binary search between these two extremes. The binary search would assume that the cost function varies monotonically and has a single point of inflexion. In other words, the cost should first reduce and then increase with increase in the number of samples. To verify this, we should plot the cost function (as well as $\measureOverlap$, $\measureExterior$ and $\measureAdjacency$) for some important elements and study it. The following elements should suffice - {\em leaf}, {\em semiquaver}, {\em treble clef}, {\em plus} and {\em S-shape}.
\fi

\nothing{
\begin{figure*}[h]
\begin{center}
\includegraphics[scale=0.64]{../resources/figure4.pdf}
\caption{Plots for $\measureOverlap$, $\measureExterior$, $\measureAdjacency$ and $\measureAsymmetry$ for various elements: x-axis represents the number of samples and y-axis shows the values of measures.}
\label{fig:figure4}
\end{center}
\end{figure*}

\begin{figure*}[h]
\begin{center}
\includegraphics[scale=0.64]{../resources/figure5.pdf}
\caption{Plots for $\measureOverlap$, $\measureExterior$, $\measureAdjacency$ and $\measureAsymmetry$ for various elements: x-axis represents the number of samples and y-axis shows the {\em band normalized} values of measures.}
\label{fig:figure5}
\end{center}
\end{figure*}
}

\note{
\begin{figure*}[!htbp]
\begin{center}
\caption{Results with {\em band normalization} - left side is the input elements while right side is the output sampling}
\label{fig:results_with_normalization}
\end{center}
\end{figure*}

\figref{fig:results_no_normalization} shows the results of our method over a varied collection of elements.
Note that the output honors the criterion\nothing{ defined in section~\ref{sec:Requirements}}.
}

\paragraph{Band normalization}

\note{
Most results in \figref{fig:results_no_normalization} exhibit close to desired sampling with our method.
However, notice the 'S' shaped element at the top-left. Instead of having a single row of samples, it got two rows running side-by-side. Similarly, the 'quaver' shaped element in the middle has multiple side-by-side elements in its left leg.
}
\iffalse
To understand this problem, let us observe how different measures ($\measureOverlap$, $\measureExterior$, $\measureAdjacency$ and $\measureAsymmetry$) vary for different elements.
Figure~\ref{fig:figure4} shows the plots for the 'S' element, the 'spiral' element and the 'X' shaped element (from the middle row of figure~\ref{fig:results_no_normalization}). Notice that in the three plots, the range of values of measures is different. 
\else
Different measures ($\measureOverlap$, $\measureExterior$, $\measureAdjacency$ and $\measureAsymmetry$) can have different range of values.
\fi
Specifically, $\measureAdjacency$ tends to have lower absolute values than others.
On the other hand, the weights ($\weightOverlap$, $\weightExterior$, $\weightAdjacency$ and $\weightAsymmetry$) associated with these measures are defined to establish their relative importance and they are going to behave correctly only when the values of measures are on an equivalent scale.
Thus, we {\em band normalize} all the measures before computing the final cost.
For {\em band normalization}, we divide each measure by its average value across all runs from $\sampleCountMin$ to $\sampleCountMax$.
\nothing{
Figure~\ref{fig:figure5} shows the plots post {\em band normalization}.
}
\note{
The final sampling after {\em band normalization} is shown in \figref{fig:results_with_normalization}.
Note that the problems observed earlier have disappeared now.
}
We perform {\em band normalization} separately from normalizing all measures to $[0 \; 1]$ range\nothing{ (as described in section~\ref{sec:Our Method})}.
However, it is possible to combine both these in a single step.

\figref{fig:comparison_pw_our} compares \cite{Phogat:2019:SVG} with our refined SEC method.

\section{Concluding Remarks}

We describe simple methods to represent given shapes with sample spheres.
We have evaluated our methods for 2D vector graphics applications including element synthesis \cite{Hsu:2020:AEF} and shape warping \cite{Phogat:2019:SVG}, and plan to explore applications in 3D modeling.
We also plan to perform more comprehensive analysis and comparisons against other methods.

\begin{acks}

We would like to thank the anonymous reviewers for their valuable feedback.

\end{acks}

\ifdefined\sigchi
\balance{}
\fi

\ifdefined\sigchi
\bibliographystyle{sigchi}
\else
\bibliographystyle{ACM-Reference-Format}
\fi

\bibliography{filtered}

\ifthenelse{\equal{\final}{0}}
{
\clearpage
\pagenumbering{roman}
\section{SIGGRAPH Information System}
\label{sec:sis}

\paragraph{Executive Summary (Maximum 50 words)}

We present methods to approximate given shapes with sample spheres to balance between accuracy, simplicity, efficiency, quality, and user controllability.

\section{Video}
\label{sec:video}

\subsection{Title}

\audio[title]
{
Hi, I am Lee E.
I will describe simple methods to represent shapes with sample spheres, with my co-authors Arjun, Shally, and Tarun.
}

\paragraph{Background}

\audio[chi]
{
I have a paper in KAI 2020 about synthesizing element patterns.
Users can specify a sparse set of elements, and the system can automatically fill-in the rest of the output domain.
}

\audio[representation]
{
A core part of the method is to represent element shapes with sample spheres.
However, in the KAI paper, the samples are manually placed by the users, which might not be desirable for real applications.
}

\audio[representation_auto]
{
Thus, our goal is to automatically generate sample spheres from given shapes, so that users can directly load different graphical elements for pattern synthesis.
}

\audio[application]
{
This problem has a long history in computer graphics, with applications in simulation, rendering, and modeling.
}

\audio[desiderata]
{
In addition to automatic computation, the samples should accurately represent the shapes.
The computation should be efficient and simple to implement.
The number of samples should be small, and controllable depending on the intended applications, such as whether the shape is rigid or deformable.
}

\subsection{Method}

\audio[two]
{
Now let's talk about our two methods, with different characteristics in representing shapes with spheres.
}

\audio[mat]
{
Our first method selects polar balls along medial axis as sample spheres.
Medial axis is like the central skeleton of the shape.
Polar balls are centered at the medial axis, and their radii are set up so that each polar ball touches multiple points on the boundary of the shape.
}

\audio[mat_select]
{
We select the polar balls in decreasing sizes, since larger balls tend to better represent the shape.
}

\audio[mat_overlap]
{
We can tune the density of the balls by thresholding their overlaps, so that no two balls are closer than a given threshold.
}

\audio[sec]
{
Our second method is based on k-means smallest enclosing circles.
}

\audio[sec_points]
{
We start by placing points uniformly over the shape, and initializing a given number of sample spheres.
The sizes and locations of these spheres will be optimized iteratively.
In this toy example, we use two sample spheres visualized in red and green.
}

\audio[sec_iterate]
{
In each iteration, we assign the nearest sample for each point, and update the sample spheres as the smallest enclosing circles for each point set.
The iteration can continue until sufficient convergence or reaching a certain time limit.
}

\audio[mat_vs_sec]
{
Our two algorithms have different behaviors.
Intuitively, the MAT method fits more sample spheres within the shapes, while the SEC method generates less spheres covering the shapes.
}

\audio[sec_choice]
{
It is possible to have more samples with better fits for the SEC method to achieve properties similar to the MAT method.
Thus, that is our default choice.
}

\audio[sec_refine]
{
We can optimize the number of samples based on a set of objectives, including overlap between samples, areas outside the shape, the number of adjacent samples, and symmetry of the shape.
More details can be found in our short paper.
}

\audio[implementation]
{
We could not release our source code right now.
Fortunately, our methods are very easy to implement.
With open source code, it took me only a few hours to implement both methods.
}

\subsection{Results}

\audio[result]
{
Here are some results produced by our SEC method with optimized number of samples over a variety of shapes.
}

\audio[comparison]
{
Here is a comparison with the current automatic pin placement method for puppet warp in Adobe Illustrator.
Our method can place samples better represent the shapes.
}

\subsection{Conclusion}

\audio[conclusion]
{
Our methods could appear in a future release of Adobe products and directly benefit many users.
We plan to investigate applications of our methods in 3D, and perform more analysis and comparisons.
Thank you.
}

}
{}

\end{document}